# DOUBLY STOCHASTIC CONTINUOUS-TIME HIDDEN MARKOV APPROACH FOR ANALYZING GENOME TILING ARRAYS[1]

By W. Evan Johnson, X. Shirley Liu and Jun S. Liu

*Brigham Young University, Dana-Farber Cancer Institute and Harvard School of Public Health, and Harvard University*

Microarrays have been developed that tile the entire nonrepetitive genomes of many different organisms, allowing for the unbiased mapping of active transcription regions or protein binding sites across the entire genome. These tiling array experiments produce massive correlated data sets that have many experimental artifacts, presenting many challenges to researchers that require innovative analysis methods and efficient computational algorithms. This paper presents a doubly stochastic latent variable analysis method for transcript discovery and protein binding region localization using tiling array data. This model is unique in that it considers actual genomic distance between probes. Additionally, the model is designed to be robust to cross-hybridized and nonresponsive probes, which can often lead to false-positive results in microarray experiments. We apply our model to a transcript finding data set to illustrate the consistency of our method. Additionally, we apply our method to a spike-in experiment that can be used as a benchmark data set for researchers interested in developing and comparing future tiling array methods. The results indicate that our method is very powerful, accurate and can be used on a single sample and without control experiments, thus defraying some of the overhead cost of conducting experiments on tiling arrays.

**1. Introduction.** Commercial whole-genome tiling arrays have been developed that "tile" the entire genomes of organisms at a very high resolution. For example, Affymetrix has developed a set of 7 arrays that tile the entire human genome which on average contains one 25-mer oligonucleotide probe within every 35 base pairs in the genome. Other companies offer researchers

Received September 2008; revised January 2009.
[1]Supported in part by NIH Grants T32-CA009337 (WEJ), R01-HG004069 (XSL) and R01-GM078990 (JSL), and NSF Grant DMS-0706989 (JSL).

the option to select their own probes on the array, leading to the ability to consider genomic regions of their choice at very high resolution. With these data, biologists are able to interrogate the entire genome to find transcription factor (TF) binding sites, nucleosome occupancy, histone modifications, new transcripts, or alternative splicing events that discover new biological phenomena and confirm previous hypotheses.

Tiling array technology presents many new statistical and computational challenges to researchers. One new issue presented by tiling array experiments is that the measured intensity values from probes that map to nearby genomic regions are expected to be highly correlated, compared to traditional array analysis where the probes are typically treated as independent. To complicate analysis further, the increased genomic coverage of tiling arrays results in data that are noisier, mainly because the higher resolution reduces options for probe selection and because there is typically no preset region of interest (such as probe set in traditional microarray design). Additionally, a typical tiling Affymetrix array experiment on the whole human genome with triplicate treatment and triplicate controls can produce as many as 270 million data points, in which a researcher may only be looking for a few hundred interesting genomic sites. Therefore, tiling array analysis requires highly sensitive statistical methods that balance the logistical constraints of working with such large data sets.

Tiling array analysis is typically accomplished in two steps: *background subtraction* and *normalization* followed by genomic region localization or *peak-finding*. Background subtraction and normalization are typically applied to microarray data to filter out chip and probe background effects in the data. Researchers have proposed simultaneous normalization/background subtraction methods [Li, Meyer and Liu (2005), Johnson et al. (2006), Huber, Toedling and Steinmetz (2006), Song et al. (2007)] that can be applied to arrays individually and model individual probe behavior, attempting to filter any probe-specific bias out of the data. The reader is referred to the references above for more details on tiling array normalization and background subtraction.

Several researchers have developed methods for peak finding in tiling array analysis. Some of the first peak-finding methods were based on sliding window scan statistics estimated using a fixed number of probes [Ji et al. (2008), Gottardo et al. (2008) or by pooling probes within a genomic window of fixed size Cawley et al. (2004), Johnson et al. (2006)]. These window-based methods are developed using nonparametric or robust methodology to protect against outlying measurements at the cost of statistical power or efficiency. Additionally, windowing methods are less accurate in precisely identifying the actual start and end positions of the regions of interest [described in detail in Huber, Toedling and Steinmetz (2006)]. In addition, scan statistics based on a fixed number of probes do not take into account the fact that the probes are unequally spaced across the genome and, therefore, probes that are very distant from each other could be combined or pooled. Other peak finding algorithms do allow for more flexible peak size. However, these methods either require a user-defined peak size distribution [Keles (2006)], a user-defined peak region shape [Zheng et al. (2008)] or the specification of the expected number of peak regions [Huber, Toedling and Steinmetz (2006)].

In this paper we propose a model for peak finding in tiling array applications using a continuous-space latent Markov model. The observed probe intensity values are modeled using a two component mixture model that individually estimates the probability that each probe is *differentially hybridized* in the treatment samples compared to the control samples (or hybridized at a level above background for applications/analyses without control samples). Then, based on the probability of differential hybridization, the model considers neighborhoods of probes, looking for genomic regions with high percentages of differentially hybridized probes using a continuous-space Hidden Markov Model (HMM). This model structure naturally deals with cross-hybridized probes by allowing for a small percentage of differentially hybridized probes in the nonpeak regions, making the method very robust. Additionally, the model deals with nonresponsive probes by allowing for a percentage of the peak region probes to be nonhybridized, where previous analyses often ignore these or attempt to locate and discard these before the analysis. The continuous-space Markov assumption naturally deals with differences in probe spacing, asserting that correlation of one probe with its nearest neighbor is dependent on the distance between the midpoints of the probes, meaning that probes very close together are expected to be highly correlated and probes very far away from each other are nearly independent. We are unaware of any other peak finding method in the literature that explicitly accounts for this differential probe spacing. Furthermore, the mixture component for the hybridized probes is assumed to have a hierarchical Bayesian structure, allowing different peaks to have different means, and estimation of these means are a byproduct of the model-fitting procedure, providing a measure for the magnitude of each peak region. Finally, the model can handle most applications on tiling arrays and can easily be adapted to analyze data from experiments with or without replicate samples and with or without control samples.

**2. Data examples and preprocessing.** We applied our method on two tiling array datasets: (1) the the recent *S. cerevisiae* tiling experiment presented in David et al. (2006) and (2) a spike in study presented in conducted as part of the Human ENCODE project presented in Johnson et al. (2008). We use the first dataset to illustrate our method and also to compare with the method of Huber, Toedling and Steinmetz (2006). The second dataset is a spike-in experiment that can be used as a benchmark data set for comparing existing and future tiling array analysis methods. More details of these data sets are given below.

2.1. *Transcript finding.* The *S. cerevisiae* tiling experiment was designed for mapping actively transcribed regions in the entire yeast genome. In this experiment, total RNA was isolated from yeast cells and enriched for Poly(A) RNA. First-strand cDNA was synthesized using random primers, and then labeled and hybridized on a microarray. The array tiles both strands of each chromosome using overlapping 25-mer probes that have an average of 8 bp between probe midpoints. The experiment included three replicate hybridizations for the RNA samples and no control samples.

We mapped the microarray probes to the yeast genome using the software presented in Li et al. (2008). There were $\sim$2.8 million probes that mapped to a unique location in the genome and any probes that mapped to multiple genomic locations were discarded. We then applied the method of Johnson et al. (2006) to normalize the data and take out bias and variation in the data that can be attributed to probe and sample effects.

2.2. *ENCODE spike-in study.* The spike-in study was conducted on the Affymetrix Human ENCODE version 1.0 array which contains approximately 700K probes and tiles nearly 1% of the human genome. In the experiment, 96 clones approximately 500 bps in length were spiked into sample at $(2^n + 1)$-fold enrichment for $n = 1, \ldots, 8$ compared to genomic DNA. Some of these clones mapped to overlapping locations on the genome and a few of the clones mapped to locations that were not tiled on the array. Control samples consisted of sonicated DNA that were labeled and hybridized on the array.

There were 70 unique spike-in regions and the number of probes in each region ranged from 2 to 94 probes, with a median of 21. The size of the regions covered on the array ranged from 44 bps to 2044 bps, with a median of 465. The probes on the array are 25 bps long and the midpoints of consecutive probes are spaced at an average of 35 bps. The analysis in the sections below includes three arrays for both the treatment and control samples. We first preprocessed these samples using the standardization method of Johnson et al. (2006).

**3. Tiling array model definition.** We apply a continuous-time Markov process to model tiling array data to locate regions of protein binding or areas of active transcription. Our model definition below equates to four-state HMM, where each probe across the chromosome is assumed to fall within one of the four states. In simple terms, the model classifies each probe as having high or low intensity using a mixture model. In addition, using a Markov assumption, the model considers the status of neighboring probes. Each probe is either classified as a high or low probe in the midst of a genomic region containing mostly high or low probes (peak or nonpeak). Convolutions of probe high/low and region peak/nonpeak leads to the four states of the model. The main assumption of the model is that peak regions will have a high proportion of high-intenstity probes, whereas nonpeak regions will have a low proportion of high probes.

The inclusion of these four states has a buffering effect on outlying probes. For example, in nonpeak regions it is common to observe sporadic high-valued probes due to cross-hybrization or sequence repeats in the genome. The four-state HMM allows for the probe status of these probes to change almost freely, but considers several probes at once to determine region status. This buffers the effect of cross-hybridized and nonresponsive probes and avoids the frequent region state-switching—making the model more robust and accurate.

3.1. *Formal model definition.* To formally define our model for peak-finding, let $i = 1, \ldots, N$ index the probes from the tiling array that map to a given chromosome, where the probes are ordered by increasing genomic location $L_i$, and the distance between probes given as $d_i = L_{i+1} - L_i$. Let $\mathbf{X} = (\mathbf{X_1}, \ldots, \mathbf{X_N})$, where $\mathbf{X_i} = (X_{i1}, \ldots, X_{in_c})^T$, be the observed normalized logged intensity values from the $n_c$ control samples in the experiment, and likewise let $\mathbf{Y} = (\mathbf{Y_1}, \ldots, \mathbf{Y_N})$, where $\mathbf{Y_i} = (Y_{i1}, \ldots, Y_{in_t})^T$, be the observed normalized logged intensity values from the $n_t$ treatment samples. $X_{ij}$ is assumed to follow a Gaussian distribution with probe-specific parameters $\theta_{i0} = (\mu_i, \sigma^2)$, and henceforth denote $\varphi_{\theta_{i0}}(\mathbf{X_i}) = \prod_{j=1}^{n_c} f_{\theta_{i0}}(X_{ij})$, where $f_\theta(\cdot)$ is the density function of a Gaussian distribution with parameters $\theta_{i0}$. $Y_{ij}$ is assumed to be generated by a mixture of two Gaussian distributions. If the complementary DNA/RNA sequence for probe $i$ is not enriched in the treatment samples, $Y_{ij}$ will be assumed to follow the same distribution as $X_{ij}$, and if the complementary sequence is enriched in the sample, $Y_{ij}$ will be assumed to follow a Gaussian distribution with parameters $\theta_{i1} = (\mu_i + \delta_i, \tau^2)$, where $\delta_i$ measures difference in the abundance of the complementary sequence for probe $i$ in the treatment versus control samples. We will define $\varphi_{\theta_{ij}}(\mathbf{Y_i})$ as with $\varphi_{\theta_{ij}}(\mathbf{X_i})$ above.

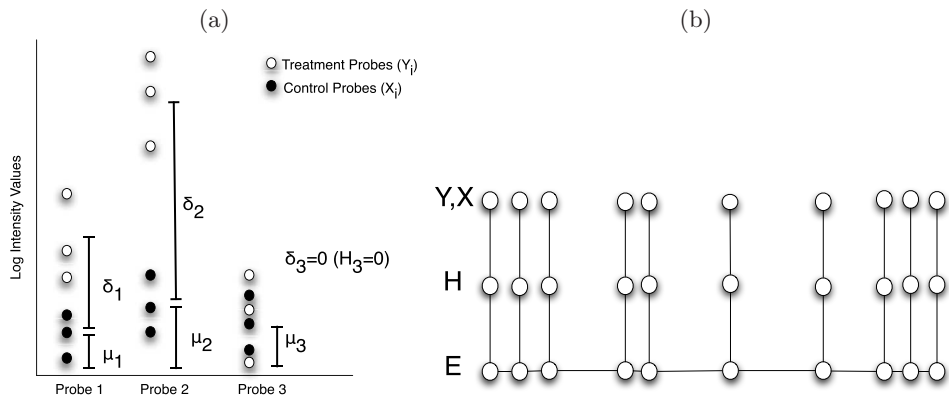

FIG. 1. (a) *Illustration of the random probe effects portion of the tiling array model.* (b) *Correlation structure of the defined tiling array model. There is one set of (three) nodes for each probe on the array. $Y, X$ represent the observed data, $H$ is a latent variable indicating the hybridization status of the probe, and $E$ is a latent continuous-time Markov chain that indicates whether the region is a peak region.*

As illustrated in Figure 1(a), we assume a hierarchical or random effects structure on $\mu_i$ and $\delta_i$, meaning that the probe-specific means and abundance levels are themselves random variables drawn from a common distribution across all probes on the array. More formally, we assume that $\mu_i \sim N(\mu, \eta^2)$ and $\delta_i \sim N(\delta, \xi^2)$, where $\mu$, $\eta^2$, $\delta$ and $\xi$ are the same for all probes on the array. For future reference, let $\theta_0 = (\mu, \eta^2)$, $\theta_1 = (\delta, \xi^2)$, and $\phi_{\theta_{i0}}(\mathbf{Y_i}) = \varphi_{\theta_{i0}}(\mathbf{Y_i})\varphi_{\theta_0}(\mu_i)$ and $\phi_{\theta_{i1}}(\mathbf{Y_i}) = \varphi_{\theta_{i1}}(\mathbf{Y_i})\varphi_{\theta_0}(\mu_i)\varphi_{\theta_1}(\delta_i)$. The model design assumes that the control samples are draws from a distribution with a mean of $\mu_i$. Including the hierarchical assumption will stabilize the estimates of $\mu_i$ and $\delta_i$ in experiments with small sample size.

We define $\mathbf{H} = (H_1, \ldots, H_N)$ as a set of unobserved indicator variables that take on values

$$H_i = \begin{cases} 1, & \text{if probe } i \text{ has mean } \mu_i + \delta_i, \\ 0, & \text{if probe } i \text{ has mean } \mu_i. \end{cases}$$

Therefore, $H_i = 0$ indicates that the observed probe signal measures probe-specific background noise and $H_i = 1$ indicates that the probe is specifically or *differentially hybridized*, or meaning that there is an enrichment of the complementary DNA/RNA sequence in the sample hybridized on the array. Henceforth, these probes will merely be referred to as *hybridized* probes. Additionally, as with the intensity values, we assume that $H_i$ is independent of $H_j$ for all $i$ and $j$, given that the region status (peak or nonpeak) is known. Additionally, we assume that intensity values are independent of each other across probes given the hybridization status of the probes. Peak regions can be identified as regions with high percentages of hybridized probes. Not all the probes in these regions are required to be hybridized, allowing for a percentage of the probes to be nonresponsive, a phenomenon commonly observed in array experiments. Additionally, the nonpeak genomic regions are allowed sporadic high probe values, accounting for cross-hybridizations which occur when an incorrect sequence binds to the probe or when the probe's complementary sequence appears in multiple locations on the genome. More formally, this equates to $H_i$ being derived from a mixture of Bernoulli distributions, with parameters $p_0$ and $p_1$ for the nonpeak and peak regions respectively. To identify peak regions, we define another set of latent indicators given by $\mathbf{E} = (E_1, \ldots, E_N)$ that take on values

$$E_i = \begin{cases} 1, & \text{if probe } i \text{ is in a peak region,} \\ 0, & \text{if probe } i \text{ is not in a peak region.} \end{cases}$$

We assume that $\mathbf{E}$ is a two-state continuous-time Markov chain with a transition matrix defined below. Convolutions of $H_i$ and $E_i$ lead to a four state HMM where transitions between hybridization states are independently determined by the individual probes and transitions between region status are governed by the Markov transition matrix.

We use the transition matrix defined in the the doubly stochastic Poisson process often referred to as a *Cox Process* [Cox (1955)]. We assume that the distance between peak regions and the size of the peak regions follow *Exponential*($\mu_0$) and *Exponential*($\mu_1$) distributions, respectively. Under this assumption, the transition matrix is given by

(1) $$\mathbf{T}(d) = \begin{pmatrix} \pi_0 + \pi_1 \exp(-kd) & \pi_1\{1 - \exp(-kd)\} \\ \pi_0\{1 - \exp(-kd)\} & \pi_1 + \pi_0 \exp(-kd) \end{pmatrix},$$

where $k = \mu_0 + \mu_1$, $\pi_0 = \mu_1/(\mu_0 + \mu_1)$, $\pi_1 = 1 - \pi_0$ and where $d$ is the genomic distance between consecutive probes. Figure 1(b) contains a schematic diagram illustrating the correlation structure of the tiling array model described here. A detailed complete data likelihood of the observed and missing data is given in the supplemental article [Johnson, Liu and Liu (2009)].

The data generating process assumed by the model, as illustrated in Figure 1, consists of partitioning all contiguously tiled genomic regions into peak and nonpeak regions using a continuous-time Markov process. Then each probe is designated as hybridized or nonhybridized with probability $p_0$ if the probe is in a nonpeak region and $p_1$ if the probe is in a peak region. Then given hybridization status, the probe is given intensity values from $\varphi_{\theta_{i0}}(\mathbf{Y_i})$ for controls and nonhybridized probes and $\varphi_{\theta_{i1}}(\mathbf{Y_i})$ for hybridized probes.

3.2. *Bayesian implementation.* We will use a fully Bayesian approach to estimate the parameters and latent states in the model. We impose the following prior distributions on the model parameters:

$$\begin{aligned} \mu &\sim N(m, s^2), & \sigma^2 &\sim \text{IG}(a_\sigma, b_\sigma), \\ \delta &\sim N(d, t^2), & \tau^2 &\sim \text{IG}(a_\tau, b_\tau), \\ p_0 &\sim \text{Beta}(a_0, b_0), & \eta^2 &\sim \text{IG}(a_\eta, b_\eta), \\ p_1 &\sim \text{Beta}(a_1, b_1), & \xi^2 &\sim \text{IG}(a_\xi, b_\xi), \\ \pi &\sim \text{Beta}(a_\pi, b_\pi), & k &\sim \text{Gamma}(a_k, b_k). \end{aligned}$$

Hyperprior values were selected based on the application; but in general, these have very little effect on the results because of the massive amounts of data that are associated with tiling arrays. These priors are conjugate for the parameters $\mu$, $\delta$, $\sigma^2$, $\tau^2$, $\eta^2$, $\xi^2$, $p_0$ and $p_1$.

3.3. *Adapting the model for nontraditional designs.* Some applications on tiling arrays require control samples, but other experiments have no natural control, for example, mapping all actively transcribed genes in the genome [Kapranov et al. (2002), David et al. (2006)]. In addition, although including replicate and control samples generally increase sensitivity and decrease false positive rates of the peak detection algorithm, recent research has shown that it is possible to achieve consistent results with a single sample analyses without replicate samples and/or controls [Johnson et al. (2006), Song et al. (2007)]. This is a very beneficial result for exploratory studies or in experiments that are very expensive to conduct. For this reason, we have adapted our model to accommodate experimental designs that do not include controls or replicates or neither.

In experiments without replicate samples and/or controls, we simplify the model by omitting the probe-specific random effects and assume that the log-intensity values $Y_i$ are derived from a mixture of Gaussian distributions with common means and variances across the probes; more specifically, the values are generated from either a $N(\mu, \sigma^2)$ or a $N(\mu + \delta, \tau^2)$ distribution. The major limitation of this model is that it does not allow for probe or region specific enrichment measurements, meaning that no natural measure of absolute enrichment is derived from the model. However, as shown in the results below, this model is able to detect true signals with high accuracy and sensitivity.

## 4. Estimation.

4.1. *An Expectation Conditional Maximization algorithm.* For efficient model estimation, we developed an Expectation Conditional/Maximization (ECM) algorithm [Meng and Rubin (1993)] to estimate the parameters of the model and then subsequently infer the latent Markov chain. The latent indicators $\mathbf{H}$ and $\mathbf{E}$ are assumed to be missing data and $\boldsymbol{\Theta} = (\mu, \delta, \sigma^2, \tau^2, \eta^2, \xi^2, p_0, p_1, \pi, k, \mu_i, \delta_i$ for $i = 1, \ldots, N)$ are treated as parameters in the model. Also let $\boldsymbol{\Theta}^{(t)}$ denote the estimates of the parameters $\boldsymbol{\Theta}$ at the $t$th iteration of the algorithm.

In the Expectation step (E-step), we let $Q(\boldsymbol{\Theta}|\boldsymbol{\Theta}^{(t)}) = E_{\boldsymbol{\Theta}^{(t)}}[\ell(\boldsymbol{\Theta}|\mathbf{y}, \mathbf{x}, \mathbf{H}, \mathbf{E})]$, where $\ell(\boldsymbol{\Theta}|\mathbf{y}, \mathbf{x}, \mathbf{H}, \mathbf{E})$ is the log of the likelihood function given in Johnson, Liu and Liu (2009). Since $H_i$ and $E_i$ are binary random variables, their expectations (and expectations of functions of $H_i$ and $E_i$) can be represented as probabilities. With this in mind, it follows that the log-likelihood is linear in $P(H_i = j|\boldsymbol{\Theta}^{(t)}), P(E_i = k|\boldsymbol{\Theta}^{(t)}), P(H_i = j, E_i = k|\boldsymbol{\Theta}^{(t)})$ and $P(E_{l-1} = j, E_l = k|\boldsymbol{\Theta}^{(t)})$ for all for all $i = 1, \ldots, N$ and $j, k = 0, 1$, and $l = 2, \ldots, N$. Therefore, we estimate these probabilities using a *forward–backward* dynamic programming algorithm and then substitute them into $Q(\boldsymbol{\Theta}|\boldsymbol{\Theta}^{(t)})$, which completes

the E-step. A detailed description of the forward–backward algorithm used here is given in the supplemental article [Johnson, Liu and Liu (2009)].

The Maximization step (CM-step) is a hybrid between a Maximization and Conditional Maximization step. It is staightforward to maximize $Q(\Theta|\Theta^{(t)})$ over the parameters $p_0$, $p_1$, $\pi$ and $k$ because they are independent of the other model parameters and $p_0$ and $p_1$ have closed-form maximizers. $Q(\Theta|\Theta^{(t)})$ is maximized over $\pi$ and $k$ using a Newton–Raphson algorithm nested within each iteration of the CM-step. The remaining parameters of the model are estimated by conditionally maximizing $Q(\Theta|\Theta^{(t)})$ iteratively over the others parameters, one at a time, while holding the other parameters at the most recent estimated value as described in Meng and Rubin (1993). A more detailed implementation of this algorithm is described in the supplemental article [Johnson, Liu and Liu (2009)].

4.2. *A Markov chain Monte Carlo approach.* We also use a Gibbs sampling approach to sample from the posterior distribution of the parameters and missing data. This approach included a *forward–backward sampling* algorithm which samples from the joint posterior of the Markov chain [see the supplemental article [Johnson, Liu and Liu (2009)] for details]. Most parameters have conjugate priors, with the exception of $\pi$ and $k$ for which we include an integrated Metropolis-type algorithm to draw from their distributions. Details for the implementation of this algorithm are given in Appendix B. In Section 5.4 we compare the ECM and MCMC algorithms.

4.3. *Region scoring.* In a tiling array analysis using the model presented here, it is of primary interest to estimate the peak region indicators $E_i$, and then rank the significant regions based on the strength of the peak. Statistical significance of peak regions can be determined using the marginal probability estimates of $E_i$, obtained using $P(E_i = 1|\Theta^{\mathbf{CMLE}})$ for the ECM algorithm (where $\Theta^{\mathbf{CMLE}}$ is the value of the parameters to which the algorithm converged) or $P(E_i = 1)$ for the MCMC method estimated by stochastic simulation.

In order to rank the regions, we utilize the marginal probabilities, and the model parameter $\delta_i$, which is designed to measure the difference in abundance of DNA/RNA fragments between the treatment and control samples, and therefore can be thought of as a measure of practical significance (under the assumption that higher values of $\delta_i$ are more interesting). Additionally, we incorporate $H_i$ in the score calculation, which basically removes nonrepsonsive probes from the region score calculation for each significant peak region. We rank the regions based on

$$Region\ Score_R = \frac{\sum_{i:i\in S_R} w_i \delta_i}{\sum_{i:i\in S_R} w_i},$$

where $S_R$ is the set of consecutive statistically significant probes that constitute region $R$, and $w_i$ is the estimate of $P(H_i = 1, E_i = 1)$ from the model. For the model that accommodates no replicates or controls, there are no $\delta_i$s in the model, so the scoring is based on the actual observed log-intensity values.

Heuristically, this scoring function given above is a compromise between the statistical significance of the individual probes in the regions, their difference in abundance between samples, and the likelihood that the probe is measuring differential hybridization or background noise. This scoring function is similar to the method used by Johnson et al. (2006), which uses a trimmed mean statistic to remove nonresponsive probes, but also has the disadvantage of removing the top scoring probes in the region. Additionally, the method presented here has the advantage of trimming specific probes whose likelihood of being nonresponsive are high as opposed to trimming a set percentage of probes in the region. In fact, the hidden **H** layer acts as a trimming agent that is optimized locally, meaning that the trimming proportion for each region can have different trimming percentages, as dictated by the probes in the region [see the supplemental article by Johnson, Liu and Liu (2009) for more detail].

## 5. Results.

5.1. *Transcript finding.* Henceforth, we denote our tiling array model described above as DSAT (Doubly Stochastic Analysis of Tiling arrays). We applied DSAT to the transcription data using the ECM algorithm with a probability cutoff of 0.9. We defined the convergence criterion for the ECM algorithm to be that the complete data log likelihood must change by less than 0.001. The algorithm converged in 1 hour 48 minutes on a single CPU of a Macintosh computer with two 3GHZ Quad-Core Intel Xeon processors.

To evaluate the performance of DSAT on this dataset, we obtaned a list of 6,604 open reading frames (ORFs) for the yeast genome (SGD, http://www.yeastgenome.org/). Of these ORFs, 6475 were tiled by probes on the array, ranging from 1 to 1810 probes per ORF, with a median 129. The number of bps tiled for each ORF ranged from 25 to 14,753, with median of 1081. Of the total base pairs called significant by DSAT, 87.6% fell on either strand within ORF boundaries [compared with 84% reported in David et al. (2006)] and 91.4% of the ORFs contained some significant expression in the DSAT regions [compared to 90% reported in David et al. (2006)]. Figure 2(a) shows an up-close view of one region from chromosome 4.

We also compared DSAT with the segmentation-based transcript discovery method from Huber, Toedling and Steinmetz (2006). This method requires the pre-specification of the number of segments, which we specified by predicting that the average segment length is 1500 bps as recommended by Huber, Toedling and Steinmetz (2006). We analyzed the data using all three replicates simultaneously (3R), the three replicates individually (R1, R2, R3) and then all three replicates but removing 90% of the probes ($3R_{10}$). We used the top scoring 1000 3R regions from each method to compare the consistency of the results when replicates and probes are removed. For DSAT, 92%, 88% and 84% of the R1, R2, R3 regions and 82% of of the $3R_{10}$ regions were in the DSAT 3R list. For the segmentation algorithm, 83%, 77% and 77% of the R1, R2, R3 regions and only 52% of of the $3R_{10}$ regions were in the segmentation 3R list. Therefore, it appears that DSAT is much more consistent, particularly when the density of the array is reduced.

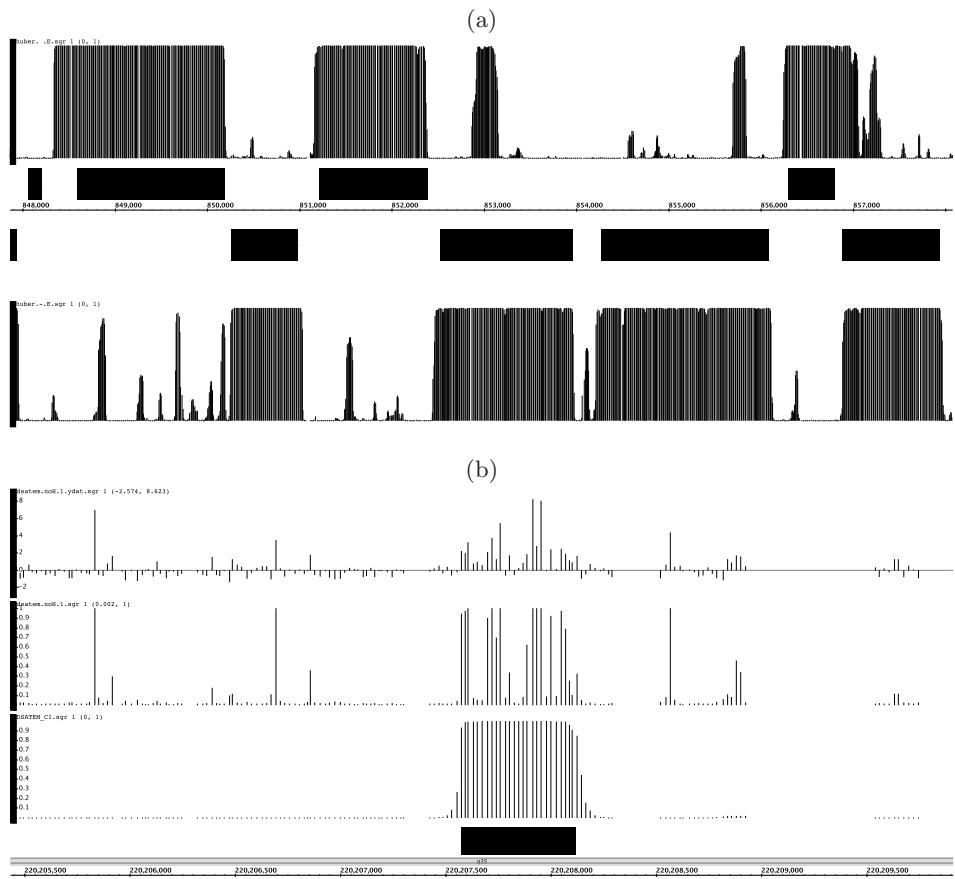

FIG. 2. *For plots* (a) *and* (b) *the x-axis represents chromosomal location. Plot* (a) *is a view from the transcript finding dataset. The top and bottom tracks are the probabilities from the model (highest ones are very near to 1) based on the data from the positive and negative strands, respectively. The second and third tracks (boxes) show the locations of annotated ORFs on the positive and negative strands, respectively. For plot* (b), *the tracks represent the probe intensity values from the S1 sample, the DSAT (ECM) probabilities after removing the hidden* **H** *layer, the DSAT (ECM) probabilities including the* **H** *layer, and the box in the fourth track indicates the true spike-in region. Notice that including the* **H** *layer buffers the noise in the data and produces a clear result.*

5.2. *Spike-in results.* We applied DSAT using the ECM and MCMC algorithms to the spike-in samples individually (S1, S2, S3), to each spike-in individually using one control (S1C1, S2C2, S3C3), three replicates with no controls (3S), three replicates with three controls (3S3C) and, for comparison with other methods, we also conducted the 3S3C analysis eliminating two-thirds of the probes on the array (used every third probe). DSAT regions were selected using a probability cutoff of 0.10 and ordered by the scoring function defined above. Normally we recommend a higher probability cutoff, but the lower cutoff was used in this case so that the number of significant DSAT regions better matched the number of regions called by the other methods (using default parameter values) in the comparisons below.

Table 1 contains the final parameter values from the S1, S1C1, 3S and 3S3C analyses for the ECM algorithm. Some of the parameter estimates of these analyses yield some interesting insight into the data. In the S1 analysis, $p_0$ indicated that an estimated 5.1% of the probes in nonspike-in regions are considered hybridized above background, while $1 - p_1$ indicates that 5.8% of the probes in the spike-in regions are not hybridized above background or are nonresponsive. Similar interpretations are appropriate for $p_0$ and $p_1$ in the analyses with controls, except that they measure the *differential hybridization* of the spike-ins versus the controls. In the single sample (with single control) featured here, we estimate that 0.6% of the probes in nonspike-in regions are differentially hybridized in the nonspike-in regions and 94.8% of the probes in the spike-in regions are differentially hybridized. In the S3C3 analysis, 0.7% of the probes in the nonspike-in regions were consistently differentially hybridized across arrays, while 97.8% of the probes in the putative spike-in regions were consistently differentially hybridized.

The values of $\pi$ and $k$ also give interesting insight into the data. $\pi$, which ranges from 0.20–0.32%, measures the approximate proportion of base pairs in the sample that are in spike-in regions. $k$ measures the approximate length (in base pairs) of the spike-in regions, ranges from 347.2–401.0 which seem to slightly underestimate the true length of $\sim 465$.

5.3. *Latent **H** layer and probe spacing.* The **H** layer is an absolutely essential element of the model. Sporadic cross-hybridized probes (or single probes in repeat regions) in nonpeak regions often have such high signals that DSAT without the **H** layer will tend to change states based on one probe and then immediately switch back. This same phenomenon also occurs

TABLE 1
*Parameter values from the ECM algorithm on the Encode Spike-in study for varying numbers of replicates and controls*

|      | $p_0$ | $p_1$ | $\mu$   | $\delta$ | $\sigma^2$ | $\eta^2$ | $\xi^2$ | $\pi$  | $k$   |
|------|-------|-------|---------|----------|------------|----------|---------|--------|-------|
| S1   | 0.051 | 0.942 | $-0.111$ | 2.25     | 0.41       | —        | 6.55    | 0.0020 | 347.2 |
| 3S   | 0.140 | 0.778 | $-0.226$ | 1.52     | 0.34       | —        | 2.61    | 0.0030 | 401.0 |
| S1C1 | 0.006 | 0.948 | 0.004   | 3.19     | 0.33       | —        | 1.56    | 0.0032 | 371.1 |
| S3C3 | 0.007 | 0.978 | 0.018   | 3.01     | 0.27       | 0.58     | 19.9    | 0.0032 | 356.0 |

S1 = The first spike-in sample without controls.
3S = The first three spike-in sample without controls.
S1C1 = The first spike-in sample with the first control.
3S3C = The first three spike-in samples with the first three controls.

in enriched/expressed regions with nonresponsive probes. To illustrate this point, we removed the **H** layer and reanalyzed the S1 sample. Figure 2(b) shows that the model without the **H** layer changes states too often. In fact, there is little difference between the normalized data (track 1) and the DSAT probabilites without the **H** layer (track 2). However, note that for the DSAT method with the **H** layer (track 3) the effect of the noisy cross-hyridized and nonresponsive probes is buffered, thus providing a clean analysis result for this region.

To show incorporating probe spacing is beneficial, we reanalyzed the S1 sample with a stationary transition matrix. There was virtually no difference in the true and false positive regions called and when probes were fairly uniformly spaced the results were indistinguishable. However, when probes were not uniformly spaced there were severe problems with the stationary transition matrix method. Figure 3 illustrates these problems. The stationary matrix carries information across large genomic distances, sometimes as much as 1000 bps, leading to less accurate region definitions. In the examples in Figure 3, the true spike-ins are both about 500 base pairs. DSAT with a stationary matrix calls regions of size 1500–2000 base pairs, while DSAT incorporating genomic distance calls regions of 500–800 base pairs. The extra thousand base pairs called by the stationary method will be very detrimental to motif searching and transcription mapping. Therefore, it is clear that incorporating genomic distance is highly beneficial. Note that methods using statistics that combine a fixed number of probes (such as BAC, TileMap and HGMM) may also suffer from this problem.

5.4. *Comparing the ECM and MCMC algorithms.* Most of the parameter estimates for the ECM and the posterior modes of the parameters from the MCMC algorithm were very similar in the analysis of the spike-in study.

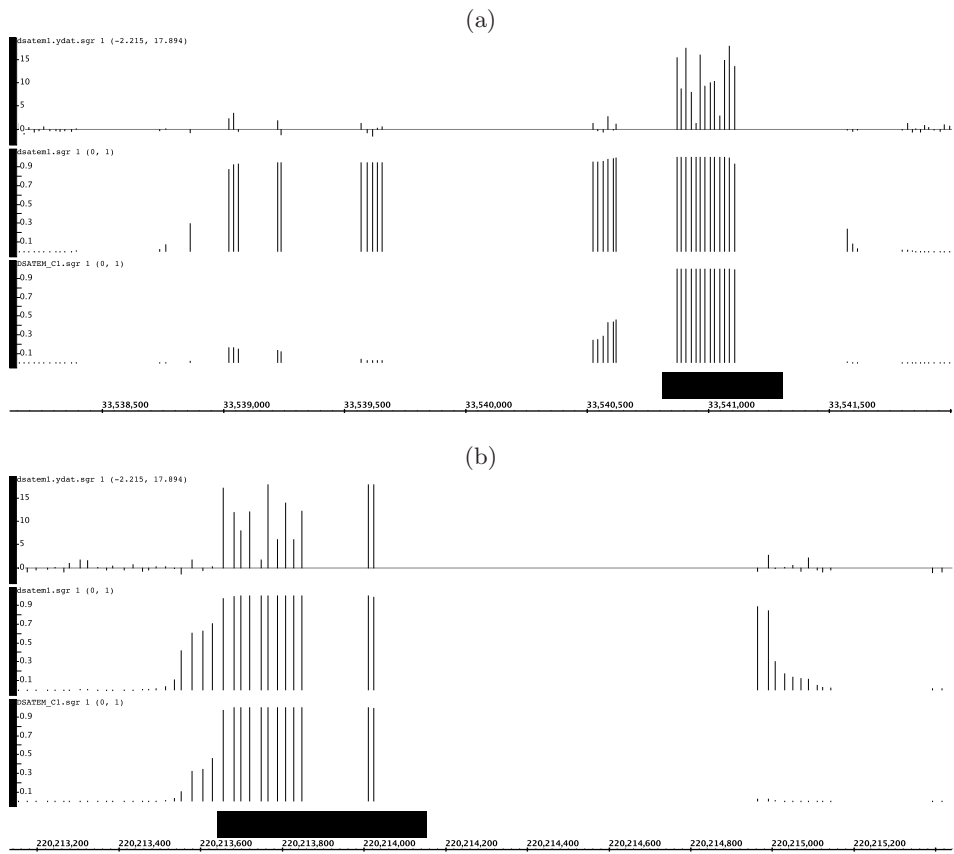

FIG. 3.    *The tracks (y-axes, top to bottom) in the plots above represent the normalized data, DSAT with a stationary transition matrix, and DSAT accounting for probe spacing. The bars at the bottom represent spike-in regions. Notice that the stationary matrix results in carrying information across large genomic distances, leading to less accurate region definitions.*

However, some of the parameter estimates and the posterior estimates of the hidden Markov chain were surprisingly different, even though they were estimated based on the same model. For example, in S1 the MCMC algorithm estimated $p_1 = 77.6\%$ (compared to 94.2% from the ECM) and the MCMC approach gave an estimate of $k = 509.7$, which is very close to the actual value (as opposed to 401.0 from the ECM method).

Based on the spike-in study, the MCMC algorithm appears to call more peak regions significant than the ECM algorithm. For example, the ECM algorithm gave between 104–118 regions in S1, S2, S3 with probability greater than 0.10, while the MCMC algorithm gave between 110–137 regions at the same cutoff (pairwise average difference is 12.7). Additionally, Figure 4(a) and (b) illustrate the increased sensitivity obtained by using the MCMC approach. In these figures, the first track is the normalized data (normalized intensity values vs. genomic position), the second track contains the ECM probabilities, the third track contains the MCMC probabilities, and the bars (if any) on the fourth track indicates the true spike-ins. Figure 4(a) shows a region with four spike-in regions in close proximity. It appears that the MCMC approach reliably identifies all four regions, whereas the ECM method only identifies three. Figure 4(c) shows the negative impact of the MCMC approach by showing a false positive region that appears to have higher significance in the MCMC vs. the ECM.

We attribute the increase in significance in the MCMC vs. the ECM to the fact that the ECM estimate for the missing peak region indicators $E_i$ is a conditional probability of $E_i$ given the parameters are fixed at $\Theta^{\mathbf{CMLE}}$, whereas the estimate from the MCMC algorithm is the posterior mean of the marginal distribution of $E_i$, integrating out $\Theta$. We noticed that a few of the parameters, in particular, the $k$ from the transition matrix, had a large effect on the results of the model, so marginalizing over the uncertainty in $k$ increased the sensitivity of the model substantially. However, in the spike-in study, this increased significance was not beneficial in terms of false discovery rate, because it appears that the ECM algorithm and MCMC algorithms find roughly the same true positive sites and so the additional sites are mostly false positives. However, in other data analyses this may not be the case, and increased significance could lead to an increase in the sensitivity of the method.

On the S1 analysis, we defined the convergence criterion for the ECM algorithm to be that the complete data log likelihood must change by less than 0.1, which turned out to be a change of ∼0.0001%. The algorithm converged in about 40 iterations which took about 25 minutes on a single CPU of a Macintosh computer with two 3GHZ Quad-Core Intel Xeon processors. The MCMC simulations took much more time; 10,500 iterations took about 24 hours to run on the same computer. Therefore, although the MCMC method was more sensitive, the ECM algorithm converged much faster and will therefore be more tractable for tiling array experiments with millions of probes.

5.5. *Comparing DSAT with other methods.* We compared the performance of DSAT with several common methods on the Affymetrix spike-in experiment. We only compare DSAT, MAT [Johnson et al. (2006)] and Tilemapv2 [Ji et al. (2008)] for single array analysis. For experiments with control samples, we compared DSAT with MAT, TileMapv2, TIMAT2 (<http://timat2.sourceforge.net/>), BAC [Gottardo et al. (2008)], HGMM [Keles (2006)], and the Affymetrix Tiling Array Suite (TAS) software, which is based on the methods used in [Cawley et al. (2004)].

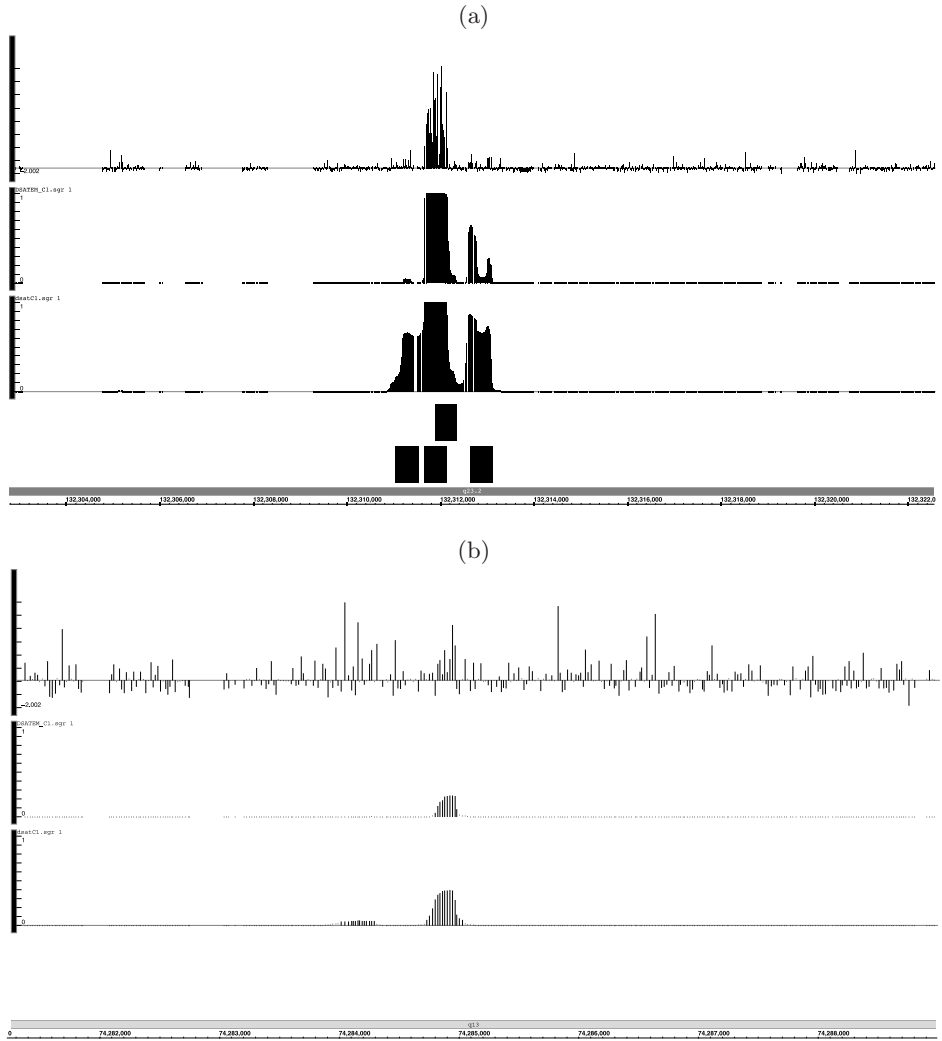

Fig. 4. *For plots* (a) *and* (b) *the x-axis represents chromosomal location. The first track represents the probe intensity values from the S1 sample, the second track represents the ECM probabilities, the third track contains the MCMC probabilities, and the boxes in the fourth track indicate the true spike-in regions.*

For BAC, we set the number of MCMC iterations to 25,000 and we still found true spike-in regions at a joint probability cutoff of 0.01. For HGMM, which requires the prespecification of peak-size distribution (based on number of probes), we inputted the actual probe distribution from the spike-in experiment. In addition, HGMM only works for positive probe values, so the normalized intensities were exponentiated, and TIMAT only works with quantile-normalized data, so we could not use the same normalization method as with the other methods. TileMap parameters were set based on the recommendation in Ji et al. (2008). All other methods and parameters were set at default values.

5.5.1. *False positive rates.* We first considered the false positive rates of the several methods on the Affymetrix spike-in experiment. Figure 5 contains a comparison between the false positive rates of the methods on the Affymetrix spike-in experiment. The $x$-axis represents the ordered significant regions for the methods and the height of the line represents the number of regions that are true spike-in regions. The plots represent the average performance for each method on the S1, S2, S3 and S1C1, S2C2, S3C3 analyses, the performance on the 3S3C analysis, and the performance on the 3S3C analysis when only one third of the probes are considered.

On the single same analyses without controls, DSAT slightly outperforms MAT, and Tilemap performs very poorly. On the single sample/single control analyses DSAT slightly outperforms MAT, TIMAT and TileMap. At around 75 regions, TileMap overcomes DSAT, although there are only 70 true regions in the data set. TAS performs very poorly on this comparison and BAC and HGMM cannot be applied to these analyses. In the 3S3C comparison, TileMap, MAT and BAC seem to work better than all the other methods. The performances of DSAT and TIMAT were similar to TileMap, MAT and BAC until about 60 regions, where DSAT and TIMAT stopped finding high-confidence regions (i.e., for DSAT the cutoff was set at 0.10). For the 3S3C analysis using every third probe, DSAT, MAT, TIMAT and TileMap performed similarly, whereas the accuracy of BAC HGMM and TAS are greatly reduced. We postulate that the poor performance here indicates that these methods may need large quantities of data for the best performance.

Also of note is that DSAT does not require the user to select parameters such as window size (genomic: MAT, TIMAT, TAS; number of probes: BAC, TileMap, HGMM), trimming percentage (MAT), or window size distribution (HGMM), which DSAT does automatically by design. Therefore, it appears that DSAT can achieve the same or better performance level with fewer user-defined parameters.

5.5.2. *Region precision.* One of the major advantages of DSAT is that it leads to more accurate and precise identification of important biological features than window-based methods. Windowing methods (either genomic or probe-based) typically identify regions that are larger (or smaller depending on the cutoff used) than the true biological features of interest. Figure 6 compares DSAT with the genomic window-based MAT algorithm, showing that DSAT can clearly identify the start/stop locations of the spike-in regions within 1-2 probes—almost independent of the probability cutoff used. MAT's identification of the start/stop is unclear and possibly cut-off dependent, meaning that one needs to find the proper window-size and cutoff combination for each data set to accurately identify the start/stop region. Methods based on combining a fixed number of probes also suffer from precision problems based on combining probes that are spaced very distant from each other on the chromosome [See the supplemental article Johnson, Liu and Liu (2009) for more details]. Precise identification of such regions are very important for follow-up analyses such as transcript mapping and motif searching (in ChIP-chip experiments).

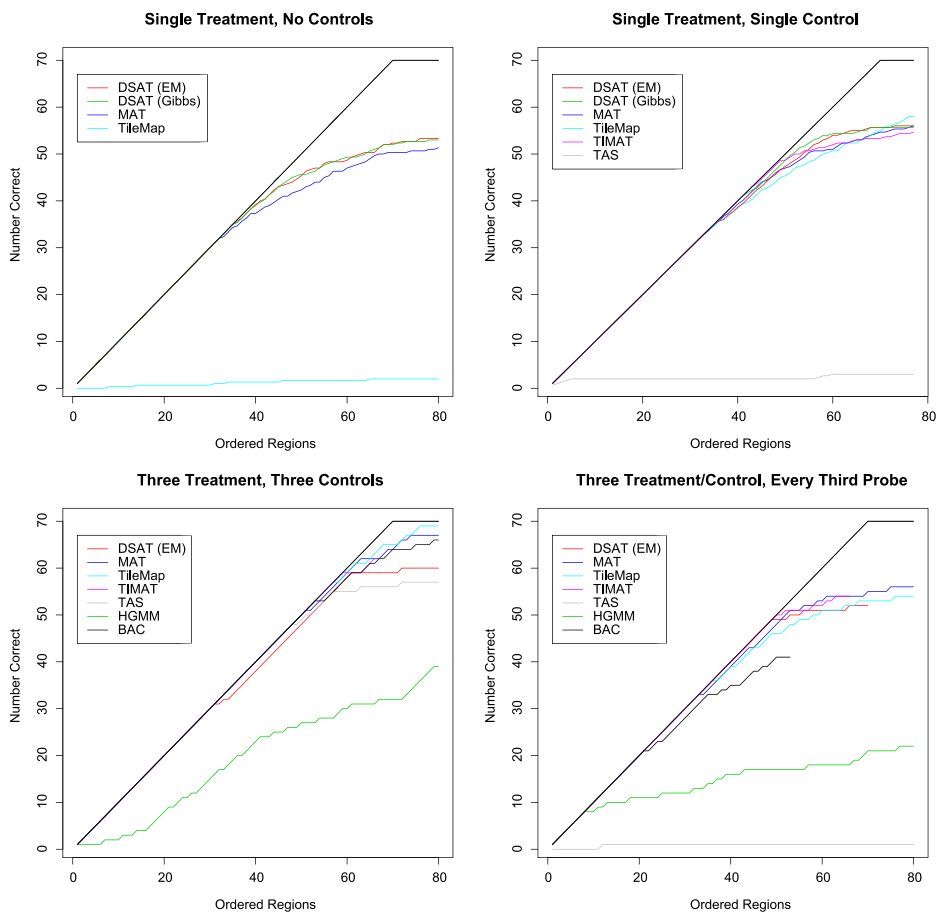

Fig. 5. *Comparison of false positive rates of various tiling array methods on the Affymetrix spike-in experiment. The x-axis represents the ordered significant regions for the methods and the height of the line represents the number of these significant regions that are true spike-in regions. DSAT appears to slightly outperform others on the single sample analysis. TileMap, MAT and BAC outperform all methods using three samples and three controls. HGMM, TAS, Tilemap and BAC seem to underperform in some of the plots on the right and bottom.*

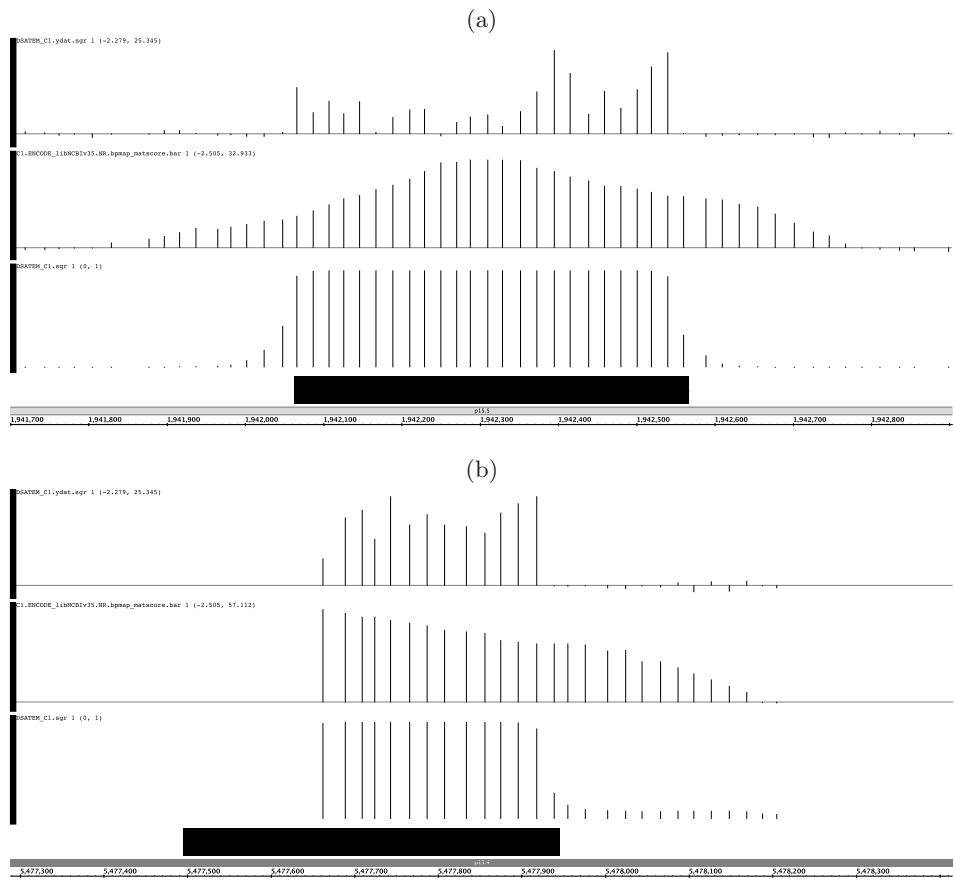

FIG. 6. *The tracks (y-axes, top to bottom) in the plots above represent the normalized data, scores from the MAT algorithm, and the DSAT (ECM) probabilities. The bars at the bottom represent spike-in regions. It is clear that DSAT can clearly identify the start/stop locations of the spike-in regions within 1-2 probes, whereas MAT's identification is unclear and possibly cut-off dependent.*

5.5.3. *Region ordering.* We also compared the spike-in region ordering from each algorithm and compared this versus the true spike-in concentration to see if the algorithms could correctly order the regions based on fold enrichment. For the regions that were correctly identified by each method, we calculated the rank (Spearman) correlation between the true spike-in level and the region score. These correlations are contained in Table 2. Based

TABLE 2
*Rank correlations between the significant regions and the true spike-in concentration for the Affymetrix spike-in experiment*

| Method | S1, S2, S3 correlations | 3S3C correlation |
|---|---|---|
| DSAT (ECM) | 0.86, 0.85, 0.87 | 0.88 |
| DSAT (MCMC) | 0.85, 0.83, 0.87 | 0.87 |
| MAT | 0.80, 0.82, 0.76 | 0.80 |
| HGMM | — | 0.76 |
| BAC | — | 0.58 |
| TAS | — | 0.51 |
| TileMap | 0.24, 0.15, 0.20 | 0.26 |
| TIMAT | — | 0.11 |

on these results, it is clear that DSAT outperforms all other methods with respect to this metric.

**6. Discussion.** Genome tiling arrays are tools that are proving to be very helpful in producing high-dimensional snapshots of biological processes. Tiling arrays are some of the most versatile and useful arrays available, as measured by the plethora of applications that are currently being conducted on them. However, tiling arrays are presenting new challenges to researchers trying to analyze and interpret the data that are produced. As the resolution of these arrays increase, the data become more correlated, variable, and contain more artifacts.

In this work we present a novel doubly stochastic latent variable model and a Bayesian analysis approach for analyzing data from many of the applications on tiling arrays. Two hidden layers of indicator variables are included in the model. One of these latent layers makes the model robust to nonresponsive and cross-hybridized probes. The other, which is assumed to follow a continuous-time Markov chain, accounts for some of the local correlation present in the data and the differential spacing between probes on the arrays, a phenomenon that has not been adequately addressed in the literature. We also allow for probe-specific random effects, allowing each probe to have its own background and hybridization distribution. The model is very flexible in that it is able to handle many applications and many different experimental designs, including experiments without replicate and/or control samples. We present two estimation approaches, the first, which utilizes an ECM algorithm, is computationally efficient and can be used on very large data sets. The second method is a Markov chain Monte Carlo method which is very sensitive because it marginalizes over all parameters in the model.

We apply our method to two datasets. The first dataset, designed for mapping active transcripts, illustrates our method in the case of a tiling array experiment without control experiments. We showed that our method handles this type of experiment well and that it outperforms a method designed specifically for this tiling array application. Additionally, we apply our method to a spike-in experiment to compare with many other tiling array methods and show that our method performs at least as well with fewer user-defined parameters. The spike-in experiment will be a very useful reference for future statisticians wishing to develop and compare new methods for tiling array applications.

**Acknowledgments.** The authors would like to thank Xihong Lin and Wei Li for helpful insights and discussions in regard to this work.

## SUPPLEMENTARY MATERIAL

**Likelihood, ECM/MCMC algorithms, and additional results and comparisons** (DOI: 10.1214/09-AOAS248SUPP; .pdf). Here we provide a detailed likelihood equation and a description of the ECM and MCMC algorithms used in this paper. In particular, we provide details on the forward–backward and forward–backward sampling algorithms used to infer the hidden Markov chain.

## REFERENCES

Cawley, S., Bekiranov, S., Ng, H. H., Kapranov, P., Sekinger, E. A., Kampa, D., Piccolboni, A., Sementchenko, V., Cheng, J., Williams, A. J., Wheeler, R., Wong, B., Drenkow, J., Yamanaka, M., Patel, S., Brubaker, S., Tammana, H., Helt, G., Struhl, K. and Gingeras, T. R. (2004). Unbiased mapping of transcription factor binding sites along human chromosomes 21 and 22 points to widespread regulation of noncoding RNAs. *Cell* **116** 499–509.

Cox, D. R. (1955). The analysis of non-Markovian stochastic processes by the inclusion of supplementary variables. *Proc. Camb. Phil. Soc.* **51** 433–441. MR0070093

David, L., Huber, W., Granovskaia, M., Toedling, J., Palm, C. J., Bofkin, L., Jones, T., Davis, R. W. and Steinmetz, L. M. (2006). A high-resolution map of transcription in the yeast genome. *Proc. Natl. Acad. Sci. USA* **103** 5320–5325.

Gottardo, R., Li, W., Johnson, W. E. and Liu, X. S. (2008). A flexible and powerful Bayesian hierarchical model for ChIP-chip experiments. *Biometrics* **64** 468–478.

Huber, W., Toedling, J. and Steinmetz, L. (2006). Transcript mapping with high-density oligonu- cleotide tiling arrays. *Bioinformatics* **22** 1963–1970.

Ji, H., Jiang, H., Ma, W., Johnson, D. S., Myers, R. M. and Wong, W. H. (2008). An integrated software system for analyzing ChIP-chip and ChIP-seq data. *Nature Biotechnology* **26** 1293–1300.

Johnson, W. E., Li, W., Meyer, C. A., Gottardo, R., Carroll, J. S., Brown, M. and Liu, X. S. (2006). Model-based analysis of tiling-arrays for ChIP-chip. *Proc. Natl. Acad. Sci. USA* **103** 12457–12462.

Johnson, D. S., Li, W., Gordon, D. B., Bhattacharjee, A., Curry, B., Ghosh, J., Brizuela, L., Carroll, J. S., Brown, M., Flicek, P., Koch, C. M., Dunham, I., Bieda, M., Xu, X., Farnham, P. J., Kapranov, P., Nix, D. A., Gingeras, T. R.,

Zhang, X., Holster, H., Jiang, N., Green, R. D., Song, J. S., McCuine, S. A., Anton, E., Nguyen, L., Trinklein, N. D., Ye, Z., Ching, K., Hawkins, D., Ren, B., Scacheri, P. C., Rozowsky, J., Karpikov, A., Euskirchen, G., Weissman, S., Gerstein, M., Snyder, M., Yang, A., Moqtaderi, Z., Hirsch, H., Shulha, H. P., Fu, Y., Weng, Z., Struhl, K., Myers, R. M., Lieb, J. D. and Liu, X. S. (2008). Systematic evaluation of variability in ChIP-chip experiments using predefined DNA targets. *Genome Res.* **18** 393–403.

Johnson W. E., Liu, X. S. and Liu, J. S. (2009). Supplement to "Doubly-stochastic continuous-time hidden Markov approach for analyzing genome tiling arrays." DOI: 10.1214/09-AOAS248SUPP.

Kapranov, P., Cawley, S. E., Drenkow, J., Bekiranov, S., Strausberg, R. L., Fodor, S. P. and Gingeras, T. R. (2002). Large-scale transcriptional activity in chromosomes 21 and 22. *Science* **296** 916–919.

Keles, S. (2006). Mixture modeling for genome-wide localization of transcription factors. *Biometrics* **63** 10–21. MR2345570

Li, W., Carroll, J., Brown, M. and Liu, X. (2008). xMAN: Extreme MApping of OligoNu-cleotides. *BMC Bioinformatics* **9** (Suppl 1) S20.

Li, W., Meyer, C. A. and Liu, X. S. (2005). A hidden Markov model for analyzing ChIP-chip experiments on genome tiling arrays and its application to p53 binding sequences. *Bioinformatics* **21** (Suppl 1) i274–i282.

Meng, X. L. and Rubin, D. B. (1993). Maximum likelihood estimation via the ECM algorithm: A general framework. *Biometrika* **80** 267–278. MR1243503

Song, J. S., Johnson, W. E., Zhu, X., Zhang, X., Li, W., Manrai, A. K., Liu, J. S., Chen, R. and Liu, X. S. (2007). Model-based analysis of 2-color arrays (MA2C). *Genome Biology* **8** R178.

Zheng, M., Barrera, L. O., Ren, B., Wu, Y. N. (2008). ChIP-chip: Data, model, and analysis. *Biometrics* **63** 787–796. MR2395716

W. E. Johnson  
Department of Statistics  
Brigham Young University  
223 TMCB  
Provo, Utah 84602  
USA  
E-mail: evan@stat.byu.edu

X. S. Liu  
Department of Biostatistics  
  and Computational Biology  
Dana-Farber Cancer Institute  
Harvard School of Public Health  
375 Longwood Ave LW641  
Boston, Massachusetts 02115  
USA  
E-mail: jliu@stat.harvard.edu

J. S. Liu  
Department of Statistics  
Harvard University  
1 Oxford St.  
Cambridge, Massachusetts 02138  
USA  
E-mail: jliu@stat.harvard.edu